\documentclass[fleqn,twoside]{article}
\usepackage[headings]{espcrc2}
\usepackage{xspace}
\readRCS
$Id: espcrc2.tex,v 1.2 2004/02/24 11:22:11 spepping Exp $
\usepackage{graphicx}
\usepackage[figuresright]{rotating}
\usepackage{relsize}
\def\babar{\mbox{\slshape B\kern-0.1em{\smaller A}\kern-0.1em
    B\kern-0.1em{\smaller A\kern-0.2em R}}}

\newcommand{\AmS}{{\protect\the\textfont2
  A\kern-.1667em\lower.5ex\hbox{M}\kern-.125emS}}

\title{Charm and Charmonium Spectroscopy}

\author{B. Aa. Petersen\address{Physics Department, 
        Stanford University, \\ 
        382 Via Pueblo Mall, Stanford, CA 94305-4060, USA} (From the \babar\ Collaboration)%
}
       
\runtitle{Charm and Charmonium Spectroscoy}
\runauthor{B. Aa. Petersen}

\def\piz   {\ensuremath{\pi^0}\xspace}

\def\pip   {\ensuremath{\pi^+}\xspace}
\def\pim   {\ensuremath{\pi^-}\xspace}
\def\pipi  {\ensuremath{\pi^+\pi^-}\xspace}
\def\pipm  {\ensuremath{\pi^\pm}\xspace}

\def\Kbar  {\kern 0.2em\overline{\kern -0.2em K}{}\xspace}

\def\Kz    {\ensuremath{K^0}\xspace}
\def\Kzb   {\ensuremath{\Kbar^0}\xspace}
\def\KzKzb {\ensuremath{\Kz \kern -0.16em \Kzb}\xspace}
\def\Kp    {\ensuremath{K^+}\xspace}
\def\Km    {\ensuremath{K^-}\xspace}
\def\Kpm   {\ensuremath{K^\pm}\xspace}

\def\KpKm  {\ensuremath{\Kp \kern -0.16em \Km}\xspace}
\def\KS    {\ensuremath{K^0_{\scriptscriptstyle S}}\xspace}

\def\B       {\ensuremath{B}\xspace}
\def\Bbar    {\kern 0.18em\overline{\kern -0.18em B}{}\xspace}
\def\BB      {\ensuremath{B\Bbar}\xspace} 
\def\Bu      {\ensuremath{B^+}\xspace}
\def\Bp      {\ensuremath{\Bu}\xspace}
\def\Bpm     {\ensuremath{B^\pm}\xspace}
\def\BR         {{\ensuremath{\cal B}\xspace}}
\newcommand{\gev}{\ensuremath{\mathrm{\,Ge\kern -0.1em V}}\xspace}
\newcommand{\mev}{\ensuremath{\mathrm{\,Me\kern -0.1em V}}\xspace}
\newcommand{\gevc}{\ensuremath{{\mathrm{\,Ge\kern -0.1em V\!/}c}}\xspace}
\newcommand{\mevc}{\ensuremath{{\mathrm{\,Me\kern -0.1em V\!/}c}}\xspace}
\newcommand{\gevcc}{\ensuremath{{\mathrm{\,Ge\kern -0.1em V\!/}c^2}}\xspace}
\newcommand{\mevcc}{\ensuremath{{\mathrm{\,Me\kern -0.1em V\!/}c^2}}\xspace}
\def\proton      {\ensuremath{p}\xspace}
\def\antiproton  {\ensuremath{\overline p}\xspace}

\def\ccbar {\ensuremath{c\overline c}\xspace}
\def\Dbar    {\kern 0.2em\overline{\kern -0.2em D}{}\xspace}

\def\Dz      {\ensuremath{D^0}\xspace}
\def\Dzb     {\ensuremath{\Dbar^0}\xspace}
\def\DzDzb   {\ensuremath{\Dz {\kern -0.16em \Dzb}}\xspace}
\def\Dp      {\ensuremath{D^+}\xspace}
\def\Dm      {\ensuremath{D^-}\xspace}
\def\Dpm     {\ensuremath{D^\pm}\xspace}

\def\DpDm    {\ensuremath{\Dp {\kern -0.16em \Dm}}\xspace}

\def\Dstarzb {\ensuremath{\Dbar^{*0}}\xspace}

\def\Dstarpm {\ensuremath{D^{*\pm}}\xspace}

\def\Ds      {\ensuremath{D^+_s}\xspace}

\def\Dss     {\ensuremath{D^{*+}_s}\xspace}
\newcommand{\jprlBase}       {Phys.\ Rev.\ Lett.\xspace}
\newcommand{\jprBase}        {Phys.\ Rev.\xspace}
\newcommand{\jplBase}        {Phys.\ Lett.\xspace}
\newcommand{\nimBaseA}       {Nucl.\ Instr.\ Methods Phys.\ Res., Sect.\ A\xspace}
\newcommand{\plb}       [1]  {\jplBase\ B~{\bf #1}}
\newcommand{\jprl}      [1]  {\jprlBase\ {\bf #1}}
\newcommand{\jprd}      [1]  {\jprBase\ D~{\bf #1}}
\newcommand{\nima}      [1]  {\nimBaseA~{\bf #1}}
\def\jpsi     {\ensuremath{{J\mskip -3mu/\mskip -2mu\psi\mskip 2mu}}\xspace}

\mathchardef\Upsilon="7107
\def\Y#1S{\ensuremath{\Upsilon{(#1S)}}\xspace}

\def\epem       {\ensuremath{e^+e^-}\xspace}

\newcommand{\DsR}[2]{{\ensuremath{D_{s{#1}}(#2)^+}}\xspace}
\def\DsJ   {\ensuremath{D^{+}_{sJ}}\xspace}
\def\DsTJ   {\ensuremath{D^{*}_{sJ}(2317)^+}\xspace}

\begin{document}

\begin{abstract}
Recent experimental results in charm and charmonium spectroscopy are reviewed.
\vspace{1pc}
\end{abstract}

\maketitle

\section{INTRODUCTION}

The last few years have seen a revival of interest in charm
spectroscopy with more than a dozen new states being reported and
hundreds of new theoretical investigations being published. The advent
of the B-factories \cite{BABAR,BELLE}, with their large, charm-rich
data samples, has proven crucial to the discovery and investigation of
new charm hadron states, but other experiments have confirmed and
complemented the B-factory observations. Much interest has been
generated by several new states that do not appear to be easily
incorporated in the conventional picture of charm and charmonium
mesons.  Here, the focus is on the latest experimental results in
charm spectroscopy and the determination of the nature of the recently
discovered states.

\section{CHARM MESONS}

Before the start of the B-factories, only four charm-strange mesons
had been observed: the S-wave states \Ds and \Dss with $J^P=0^-$ and
$1^-$, and two P-wave states, \DsR{1}{2536} and \DsR{2}{2573}, assigned
to be $1^+$ and $2^+$ states. Two additional P-wave states, $0^+$ and
$1^+$, were expected, but predicted by potential models to have a
large decay width to a non-strange charm meson and a kaon
\cite{godfreyIsgur,pierro}. The observations of two narrow charm-strange
mesons below the $DK$ threshold, the \DsTJ \cite{babar2317} and
\DsR{J}{2460} \cite{cleo2460}, by the \babar\ and CLEO
collaborations have led to much speculation whether these are the
missing $0^+$ and $1^+$ states or perhaps new
types of particles \cite{DsJModelsNew}.

Several experimental studies of the \DsTJ and \DsR{J}{2460} have been
performed recently to understand their nature. \babar\ has performed a
comprehensive study \cite{DsJBR} of the possible decay to a \Ds meson
and up to two $\gamma$ or $\piz$ particles. The \DsTJ is only observed
in decay mode $\Ds\piz$, while the \DsR{J}{2460} is observed in
three modes: $\Dss\piz$, $\Ds\pipi$ and $\Ds\gamma$. None of these
decays are in disagreement with the assignment of $0^+$ and $1^+$ for
the two states. No other decay modes are observed and limits on the
branching ratios are measured \cite{DsJBR}. Isospin partners for the
\DsTJ are searched for in decays to $\Ds\pipm$, but none are observed
\cite{DsJBR}.

A separate study by \babar\ measures the absolute branching fractions
of \DsR{J}{2460} decays \cite{absDsJ}. The analysis selects \BB events
where one \B meson is fully reconstructed and used to determine the
rest frame of the second \B meson. An additional \Dpm, \Dz or \Dstarpm
meson from the second \B is reconstructed and its recoil mass spectrum
studied. A signal is found for \DsR{J}{2460} and is used to measure
$\BR(\B\to D^{(*)}D_{sJ}(2460))$. This is combined with previous
branching fraction product measurements \cite{DsJProdBF} to obtain
absolute branching fractions for \DsR{J}{2460} decays. The branching
fractions for the three observed modes add up to $(76\pm20)\%$,
indicating that most of the \DsR{J}{2460} decay modes have been
observed.

\B meson decays have also been used to determine the spin of the two
new \DsJ states. In the decays $\B\to D\DsTJ,
\DsTJ\to\Ds\piz$ \cite{BelleDsJSpin} and $\B\to
D\DsR{J}{2460}, \DsR{J}{2460}\to\Ds\gamma$
\cite{BelleDsJSpin,DsJProdBF}, the angular distribution of the \Ds
meson with respect to the $D$ meson in the \DsJ rest frame is measured.
This distribution is consistent with flat for the \DsTJ and quadratic for the
\DsR{J}{2460}, establishing that these states have spin 0 and 1,
respectively. The \DsTJ therefore has to be a $0^+$ state,
while for \DsR{J}{2460} the possibility of being a $1^-$ state can be
excluded by a similar angular analysis of $\DsR{J}{2460}\to\Dss\piz$
decays. This result indicates that the two \DsJ states are regular
charm-strange mesons, but does not explain why the potential models
underestimate their mass.

The SELEX collaboration has reported \cite{SelexDsJ26} the possible
existence of a new, narrow charm-strange meson decaying to $\Ds\eta$
and $\Dz\Kp$ with a mass of about 2632\mevcc. This state,
\DsR{J}{2632}, is not observed by FOCUS \cite{FocusDsJ26}, Belle
\cite{BelleDsJ26} or \babar\ \cite{BabarDsJ26} even though all three
have significantly larger reconstructed samples of the nearby
$\DsR{2}{2573}\to\Dz\Kp$ decays. This appears to exclude the existence
of this state.

Further studies of the $\Dz\Kp$ mass spectrum by \babar\
\cite{BabarDsj2860} instead show a peaking structure around
2.86\gevcc. The peak is visible using either $\Dz\to\Km\pip$ or
$\Dz\to\Km\pip\piz$ decays and the same peak is also seen in the $\Dp\KS$
mass spectrum. No signal is seen in simulated events or with $\Dz$
mass sideband events. Nor is the peak due to pions misidentified as
kaons. A combined fit to the $\Dz\Kp$ and $\Dp\KS$ mass spectrums
gives a mass of $2856.6\pm1.5\pm5.0\mevcc$ and a width of $48\pm
7\pm10 \mev$. The spin-parity of this state has not yet been
established, though it has already been speculated that it could be a
$J^P=3^-$ D-wave state \cite{DsJ2860Theory1} or a radially excitation of
the \DsTJ \cite{DsJ2860Theory2}.

A broad structure is also observed just below 2.7\gevcc. A fit with an
additional Breit-Wigner gives a mass of $2688\pm4\pm3\mevcc$ and a
width of $112\pm 7\pm36\mev$ for this structure. However, the fit is
not particularly good and there is a hint of structure in the same region when using events from a \Dz mass sideband. \babar\ is therefore not able to established whether
this is a new state\footnote{After this talk was given, Belle
reported \cite{BelleDsJ27} a possible \Dz\Kp state in
$\Bp\to\Dz\Dzb\Kp$ decays with a mass of $\sim 2715\mevcc$.  This
might be the same state.}.

\section{CHARM BARYONS}

Charm baryons with two light quarks provide an even richer particle
spectrum than charm mesons. Of the states without internal orbital
angular momentum ($L=0$), all nine $J^P=\frac{1}{2}^+$ states have
been known for years \cite{PDG2006}, while observation of the last of
the six $J^P=\frac{3}{2}^+$ states, the $\Omega_c^{*}$, has been
reported very recently \cite{OmegaCstar}.  Several possibly orbitally
excited charm baryons have already been observed, the latest of which
are summarized below.

One new charm baryon has been found \cite{babar2940} to decay to
$\Dz\proton$. A fit to a Breit-Wigner shape folded with experimental
resolution yields a mass of $2939.8\pm1.3\pm1.0\mevcc$ and a width of
$17.5\pm 5.2\pm 5.9\mev$. A second, larger peak is fit with a mass of
$2881.9\pm0.1\pm0.5\mevcc$ and a width of $5.8\pm 1.5\pm1.1\mev$ and
is identified as the $\Lambda_c(2880)^+$, previously discovered in decays
to $\Lambda_c^+\pipi$ decays \cite{cleoLc2880}. These
are the first charm baryons observed to decay to a charm meson and a
charmless baryon. The new state is identified as the isospin scalar
$\Lambda_c(2940)^+$, due to the absence of an isospin partner in the
$\Dp\proton$ mass spectrum. The existance of the $\Lambda_c(2940)^+$
has been confirmed by Belle \cite{Belle2940}, who observed it in the
final state $\Sigma_c(2455)^{++/0}\pi^\mp$.  Its spin-parity has yet
to be determined.

Besides single charm baryons, double charm baryons are expected to
exist with a mass between 3.5 and 3.8\gevcc for the lightest states
\cite{ccBaryonTheory}. The only evidence for their existance comes
from SELEX, which reports an excess of events in $\Lambda_c^+\Km\pip$
\cite{Selexcc1} and $\Dp\antiproton\Kp$ \cite{Selexcc2} at a mass of
$3519\pm1\mevcc$. No evidence of this state is found by FOCUS
\cite{Focuscc}, \babar\ \cite{Babarcc} or Belle \cite{Bellecc}, even
though they have $O(10)$ (FOCUS) and $O(100)$ (\babar, Belle) more
reconstructed charm baryons than SELEX. 

In the same $\Lambda_c^+\Km\pip$ mass spectrum used to search for
double charm baryons by Belle, two new regular $\Xi_c$ states are
observed with masses of $2978.5\pm2.1\pm2.0\mevcc$ and
$3076.7.5\pm0.9\pm0.5\mevcc$ and widths of $43.5\pm 7.5\pm7.0\mev$ and
$6.2\pm 1.2\pm0.8\mev$ \cite{Bellecc}.  The non-zero widths preclude
these two states, $\Xi_c(2980)^+$ and $\Xi_c(3077)^+$, from being
double-charm baryons. The existence of these two states has since been
confirmed by \babar\ \cite{BabarXic}.  Belle also observes the isospin
partner $\Xi_c(3077)^0$ in the decay
$\Xi_c(3077)^0\to\Lambda_c^+\KS\pim$.

\section{CHARMONIUM-LIKE STATES}

Many new charmonium or charmonium-like states have been found in
recent years. Several of these have received particular attention,
since they do not seem to fit in the standard \ccbar spectroscopy
picture. The most studied of the new states is the $X(3872)$, which was
first discovered by Belle \cite{BelleXdiscover} in $\Bpm\to\Kpm X(3872),
X(3872)\to\jpsi\pipi$ decays. The observation has since been confirmed
by several experiments \cite{CDFX,D0X,BabarX}. The latest measurements
of this decay \cite{BelleXMass,BabarXMass} yield a mass
$3871.2\pm0.6\mevcc$ and an upper limit on the width of
$\Gamma<2.3\mev$ at 90\% CL. 

Several other decay modes of the $X(3872)$ have been observed. Both
Belle and \babar\ have measured the radiative decay
$X(3872)\to\gamma\jpsi$ \cite{BelleRadiative,BabarRadiative} with a
branching ratio of $(19\pm7)\%$ relative to the
$X(3872)\to\jpsi\pipi$ decay mode. The observation of this mode
implies that the $X(3872)$ has even charge-conjugation parity. The
rate is much lower than what one would expect if the $X(3872)$
was a $\chi'_{c}$ state \cite{SwansonRadiative}.  Belle also
observes the decay $X(3872)\to\Dzb\Dz\piz$ \cite{BelleDDpi}. This
decay mode appears to be dominant as it is measured to have a branching
fraction $9.7\pm3.4$ times higher than the $X(3872)\to\jpsi\pipi$
decay mode.

The quantum numbers $J^{PC}$ of the $X(3872)$ have been studied using
the angular distributions of $X(3872)\to\jpsi\pipi$ decays. Both $B\to
X(3872)K$ decays \cite{BelleXMass}  and inclusively produced $X(3872)$ mesons at CDF
\cite{CDFAngular} have been used. The combined analysis exclude
all possibilities except $1^{++}$ and $2^{-+}$. The $J=2$ hypothesis
is disfavored by the observation of the near-threshold decay
$X(3872)\to\Dzb\Dz\piz$. $J=1^{++}$ is consistent with the $X(3872)$
being a $\chi'_{c1}$ or a $\Dz\Dstarzb$ bound state
\cite{X3872Molecule1,X3872Molecule2}. The latter is currently viewed as the most
likely explanation, though tetra-quark models \cite{X3872fourquak}
have not been ruled out.

A second interesting charmonium-like state is the $Y(4260)$,
discovered by \babar\ in initial-state radiation (ISR) events,
$\epem\to\gamma_{\mathrm{ISR}}Y(4260),$ $Y(4260)\to\jpsi\pipi$
\cite{Y4260BabarDiscover}. It is a wide state with a mass of
$4259\pm8^{+2}_{-6}\mevcc$ and a width of $88\pm23^{+6}_{-4}\mev$. The
observation of $Y(4260)$ in ISR events requires that
$J^{PC}=1^{--}$. The state has been confirmed by CLEO. They performed
an energy scan and observe a large increase in the cross section of
$\epem\to\jpsi\pipi$ and $\epem\to\jpsi\piz\piz$ at
$\sqrt{s}=4260\mev$ \cite{Y4260CLEOc}. The $\epem\to\jpsi\pipi$ cross
section is consistent with that measured by \babar\ in ISR events. The
observation of $Y(4260)\to\jpsi\piz\piz$ implies that it has isospin
$I=0$. \babar\ has searched for several other possible decay modes of
the $Y(4260)$: $\proton\antiproton$ \cite{Y4260pp}, $\phi\pipi$
\cite{Y4260phipi} and $D\Dbar$ \cite{Y4260DD}. No signals are observed
and upper limits on the decay rates are set. The upper limit for the
$D\Dbar$ mode is 7.6 times $\BR(Y(4260)\to\jpsi\pipi)$ at 95\% CL,
which is much smaller than for example $\frac{\psi(3770)\to
D\Dbar}{\psi(3770)\to\jpsi\pipi}\approx 500$ \cite{psi3770}. This
suggests that the $Y(4260)$ is not a radially excited $\psi$ state,
such as $\psi(4S)$. Currently the most favored explanation is that of
a $\ccbar g$ hybrid state \cite{Y4260hybrid}, which is predicted
in this mass region.

\babar\ also searched for $Y(4260)\to\psi(2S)\pipi$ decays
\cite{Y4360}.  No signal for the $Y(4260)$ state is found, but a peak
is observed at slightly higher mass. The peak is also not consistent
with the $\psi(4415)$. A fit with a Breit-Wigner gives a
mass of $4354\pm 16\mevcc$ and a width $106\pm 19\mev$. If $Y(4260)$
is the lowest mass hybrid state, this $Y(4350)$ could be an
excitation. However, more studies of are needed to confirm
the existence of this state and understand its properties.

\section{SUMMARY}

Charm spectroscopy has seen a lot of activity in the last few
years. The $\DsTJ$ and $\DsR{J}{2460}$ mesons appear to be regular
$0^+$ and $1^+$ charm-strange mesons, but new charm mesons and baryons
have been found that need further study. In charmonium the $X(3872)$
and $Y(4260)$ have been found to be $1^{++}$ and $1^{--}$, but their
decay modes and masses suggest that these are not regular \ccbar
states.  They could be the first examples of a bound $\Dz\Dstarzb$ and a
$\ccbar g$ hybrid state, respectively. More experimental studies are
needed to confirm or refute these hypotheses.

\end{document}